# High thermal conductivity of bulk epoxy resin by bottom-up parallel-linking and strain: a molecular dynamics study

Shouhang Li[1,2,#], Xiaoxiang Yu[2,3,#], Hua Bao[1,*], Nuo Yang[2,3,*]

[1] University of Michigan-Shanghai Jiao Tong University Joint Institute, Shanghai Jiao Tong University, Shanghai 200240, P. R. China

[2] State Key Laboratory of Coal Combustion, Huazhong University of Science and Technology, Wuhan 430074, P. R. China

[3] Nano Interface Center for Energy (NICE), School of Energy and Power Engineering, Huazhong University of Science and Technology, Wuhan 430074, P. R. China

## Abstract

The ultra-low thermal conductivity (~0.3 $Wm^{-1}K^{-1}$) of amorphous epoxy resins significantly limits their applications in electronics. Conventional top-down methods e.g. electrospinning usually result in aligned structure for linear polymers thus satisfactory enhancement on thermal conductivity, but they are deficient for epoxy resin polymerized by monomers and curing agent due to completely different cross-linked network structure. Here, we proposed a bottom-up strategy, namely parallel-linking method, to increase the intrinsic thermal conductivity of bulk epoxy resin. Through equilibrium molecular dynamics simulations, we reported on a high thermal conductivity value of parallel-linked epoxy resin (PLER) as 0.80 $Wm^{-1}K^{-1}$, more than twofold higher than that of amorphous structure. Furthermore, by applying uniaxial tensile strains along the intra-chain direction, a further enhancement in thermal conductivity was obtained, reaching 6.45 $Wm^{-1}K^{-1}$. Interestingly, we also observed that the inter-chain thermal conductivities decrease with increasing strain. The single chain of epoxy resin was also investigated and, surprisingly, its thermal conductivity was boosted by 30 times through tensile strain, as high as 33.8 $Wm^{-1}K^{-1}$. Our study may provide a new insight on the design and fabrication of epoxy resins with high thermal conductivity.

[#] S.L. and X.Y. contributed equal to this work.

[*] To whom correspondence should be addressed. Email: hua.bao@sjtu.edu.cn (HB); nuo@hust.edu.cn (NY)

# Introduction

Epoxy resins are classical thermoset polymers widely used in coatings, adhesives and electronic packaging for their excellent thermal mechanical properties and high electrical resistance [1]. However, most of epoxy resins have very low thermal conductivity on the order of 0.1 $Wm^{-1}K^{-1}$ [2] owing to the existence of entanglement, voids etc. [3]. In some special situations, for example, aerospace and flexible electronics fields, heat dissipation is becoming a crucial issue and the epoxy resin with higher thermal conductivity is desirable [4]. The general strategy to increase the thermal conductivity of polymers is doping materials with high thermal conductivities, such as ceramics [5-8], metals [9], carbon nanotubes [10] and graphene [11]. Adding fillers increases the cost. More importantly, the fillers will degrade the original electrical and mechanical properties of polymers and affect the performance of the relevant products. Thus, it is necessary to enhance the intrinsic thermal conductivity of epoxy resin.

The heat transport properties of cross-linked epoxy resin (CLER) are well revealed theoretically and experimentally. Varshney *et al.* [12] calculated the thermal conductivity of CLER employing both equilibrium molecular dynamics method (EMD) and non-equilibrium molecular dynamics method (NEMD) and the thermal conductivity value is determined to be in the range of 0.30-0.31 $Wm^{-1}K^{-1}$ at 300 K. By using EMD method, Kumar *et al.* [2] found the positive temperature dependence of thermal conductivity. Kline *et al.* [13] measured the thermal conductivity of epoxy resin and found the thermal conductivity value goes from 0.23 $Wm^{-1}K^{-1}$ to 0.27 $Wm^{-1}K^{-1}$ when the temperature raises from 275 K to 375 K. The heat dissipation ability of epoxy resin is similar to other polymers and it cannot meet the stringent requirement in relevant fields.

In the past decades, researchers found that polymers with highly ordered structures have appreciable thermal conductivities. Chen and Henry [14, 15] found that the thermal conductivity of polyethylene single chain can be even divergent which is several orders of magnitude larger than that of amorphous bulk PE. The amazing discovery reveals that the intrinsic thermal conductivities of polymers have not been fully excavated. Luo [16-19] and his cooperators found that polymer nanofibers with intrinsically ordered backbones, strong backbone bonds, and strong dihedral angles could be the high thermal conductivity candidates. The aligned polymers have very high thermal conductivities is also verified by experiments. Shen *et al.* [3] found the thermal conductivity of ultra-drawn polyethylene nanofibers can be as high as ~104 $Wm^{-1}K^{-1}$ at 300K. Xu *et al.* [20] found polyethylene film consists of nanofibers with crystalline and amorphous regions owns high thermal conductivity of 62 $Wm^{-1}K^{-1}$. Cao *et al.* [21] investigated on the thermal conductivity of polyethylene nanowire arrays and they found the value of the nanowire array with the diameter of 100 nm can reaches 21.1 $Wm^{-1}K^{-1}$ at 80 °C. Singh *et al.* [22] measured the thermal conductivity of chain-oriented amorphous polythiophene and the value can be as high as ~4.4 $Wm^{-1}K^{-1}$ at room temperature.

To attain ordered structure in epoxy resin, some arisen technology [23]and materials [24] based on top-down methods were adapted to enhance thermal conductivity. Akatsuka *et al.* [25] measured the thermal conductivity of liquid-crystalline epoxy resin with macroscopic isotropic structure at 30°C and the value can be as high as 0.3 - 0.96 $Wm^{-1}K^{-1}$. The exact value depends on the type of monomer and curing agent. Correspondingly, Koda *et al.* [26] calculated the thermal conductivity of liquid crystalline epoxy resins at 300K using molecular dynamics and they got the value of 0.37

- 0.96 Wm$^{-1}$K$^{-1}$, which is consistent with the experimental results. Zeng *et al.* [27] measured the thermal conductivity of the epoxy resin fibers made by electrostatic spinning technology and they found the value can be as high as 0.8 Wm$^{-1}$K$^{-1}$. In their following work [28], they found the thermal conductivity of epoxy resin fibers increases on decreasing the fiber diameter.

However, it should be noted that the conventional epoxy resin is amorphous cross-linked network prepared by the polymerization of monomers and curing agents, which is different from those amorphous linear polymers consisting of many chains. The top-down methods mentioned above like electrospinning could be applied to straighten many unconnected chains and achieve the long-range alignment. Nevertheless, for epoxy resin, top-down methods could only construct short-range ordered structure, and also fail to avoid crosslinking and long-range disorder [28], thus form network structure giving rise to strong intra-molecule phonon scatterings in heat transport contributed by covalent bonding. This is the dominant reason why top-down methods based on cross-linking does not work well in enhancement of thermal transport in epoxy resin. Therefore, how to carefully design the linking structure with long-range alignment and better understanding of heat transport mechanism in epoxy resin remains to be explored.

In this work, to further enhance the intrinsic thermal conductivity, we propose a bottom-up linking strategy to construct parallel-linked epoxy resin (PLER). The procedure is similar to the bottom-up manufacture technology, molecular layer deposition (MLD), which can be used to make high quality polymer films [29, 30]. At the same time, the effects of morphology on the thermal conductivity of epoxy resin are investigated. We first studied the relationship between the degree of crosslinking and thermal conductivity of amorphous cross-linked network. Then we calculated the thermal conductivities of bulk PLER and the strain effects are discussed. In addition, the intrinsic thermal conductivity of PLER single chain is presented.

## Model and Simulation details

The commonly used monomer and curing agent of epoxy resin are EPON-862 (di-glycidyl ether of bisphenol F) and DETDA (diethylene toluene diamine), respectively. The corresponding chemical structures are shown in Fig. 1(a) and (b). The two end carbon atoms in EPON-862 and the two nitrogen atoms in DETDA are reaction sites and they can form a new covalent bond as shown in Fig. 1(c). Several methods have been proposed to construct crosslinking atomic model of epoxy resin network, for which the properties can be consistent with experiments [31-33]. In this work, we employed the methods described in the Ref. [33], and the relaxed cross-linked network is shown as the Fig. 1(d). In order to build ordered epoxy resin structure, we also changed the conventional crosslinking method and proposed a new parallel-linking method, which is similar to the principle of MLD. MLD is widely used to manufacture polymer films with high quality [29, 30]. The polymer films are fabricated by stacking molecules on substrates layer by layer. The schematic of MLD is shown in Fig. 1(e). For parallel-linking model, the monomer and curing agent form different molecular layers. Then one monomer reacts with one curing agent and then we could get one cross-linked segment, as shown in Fig. 1(c). Several molecular layers replicate along the chain direction (z axis) and then we could get the stacked semi-crystal structure, as shown in Fig. 1(f). In this study, there are 4 chains on x and y axes for bulk PLER, respectively.

The molecular dynamics simulations are performed at the temperature of 300K using the

Large-scale Atomic/Molecular Massively Parallel Simulator (LAMMPS) package[34]. According to the former simulations [2, 12], the consistent valence force field (CVFF) [35, 36] is accurate enough to characterize the thermal properties of epoxy resins and therefore it was employed to describe the interatomic interactions in this work. As Kumar *et al.* [2] pointed out, the simulation results can be consistent with the experiments when the long-range interactions are included. Thus, both van der Waals (vdW) force and electrostatic force are included in our simulation. The exact formula of CVFF force field is:

$$E_{total} = \sum_b K_b (r-r_0)^2 + \sum_\theta K_\theta (\theta - \theta_0)^2 + \sum_\phi K_\phi (1 + d\cos(n\phi)) + \sum_{i>j} 4\varepsilon \left[ \left(\frac{\sigma}{r_{ij}}\right)^{12} - \left(\frac{\sigma}{r_{ij}}\right)^6 \right] + \frac{1}{2} \sum_{i=1}^{N} \sum_{j=1, j\neq i}^{N} \frac{q_i q_j}{\varepsilon_0 r_{ij}}. \quad (1)$$

The terms on right hand side of Eq. (1) stand for bond force, angle force, dihedral force, vdW force and Coulomb interaction, respectively. The Lennard-Jones (LJ) potential parameters across different types of atoms were calculated by using the Lorentz-Berthlot mixing rules (i.e., $\varepsilon_{ij} = sqrt(\varepsilon_i \varepsilon_j)$, $\sigma_{ij} = (\sigma_i + \sigma_j)/2$). The cutoff distance of the 12-6 LJ potential for all structures was set to 9 Å. All potential parameters are available in Ref. [35] and [36].

The calculation of heat flux is a critical step in EMD method. The heat flux is defined as:

$$\vec{J} = \frac{1}{V} \left[ \sum_i e_i \vec{v}_i + \frac{1}{2} \sum_{i<j} \left( \vec{f}_{ij} \cdot (\vec{v}_i + \vec{v}_j) \right) \vec{x}_{ij} \right], \quad (2)$$

where $V$ is the volume of the simulation cell, $i$ and $j$ are the indexes of atoms, $e_i$ is the total energy of atom $i$, $v_i$ and $v_j$ are atom velocities of atom $i$ and $j$, $f_{ij}$ is the force between atom $i$ and $j$, $\vec{x}_{ij}$ is the relative position of atom $i$ and $j$.

We can obtain the thermal conductivity by directly integrating heat current autocorrelation function (HCACF), namely

$$\kappa = \frac{V}{3k_B T^2} \int_0^{\tau_0} \langle \vec{J}(0) \cdot \vec{J}(\tau) \rangle d\tau, \quad (3)$$

where $T$ is the system temperature, $k_B$ is the Boltzmann constant, $\tau_0$ is the integral upper limit of HCACF, which is also called correlation time. The angular bracket denotes an ensemble average.

The velocity Verlet algorithm [37] is employed to integrate equation of motion, and the time step is set as 0.25 fs. The initial structures are first minimized by standard conjugate-gradient energy-minimization methods in LAMMPS. After that, the system runs in the isothermal-isobaric ensemble (NPT) for 250 ps to relax the whole system at the given temperature and pressure. Then it runs another 250 ps in the canonical ensemble (NVT) at the given temperature 300 K. Later on, it runs in the microcanonical ensemble (NVE) for 250 ps for further relaxation. Finally, it runs at least another 2.5 ns in NVE, during which the heat current is recorded at an interval of 2.5 fs. To achieve better ensemble average, we obtained the final results on the average of at least 4 independent simulations with different initial conditions. All the aforementioned simulation parameters were carefully checked to ensure the results were converged [38].

Note that the thermal conductivities were averaged over three orthogonal directions for CLER to get better statistics as they were considered to be isotropic in nature for cross-linked network structure. For parallel-linked bulk and single chain structures, the thermal conductivity on the z axis is different from that on the other two axes due to anisotropy. For these anisotropic cases, we calculate the thermal conductivities on the three axes separately.

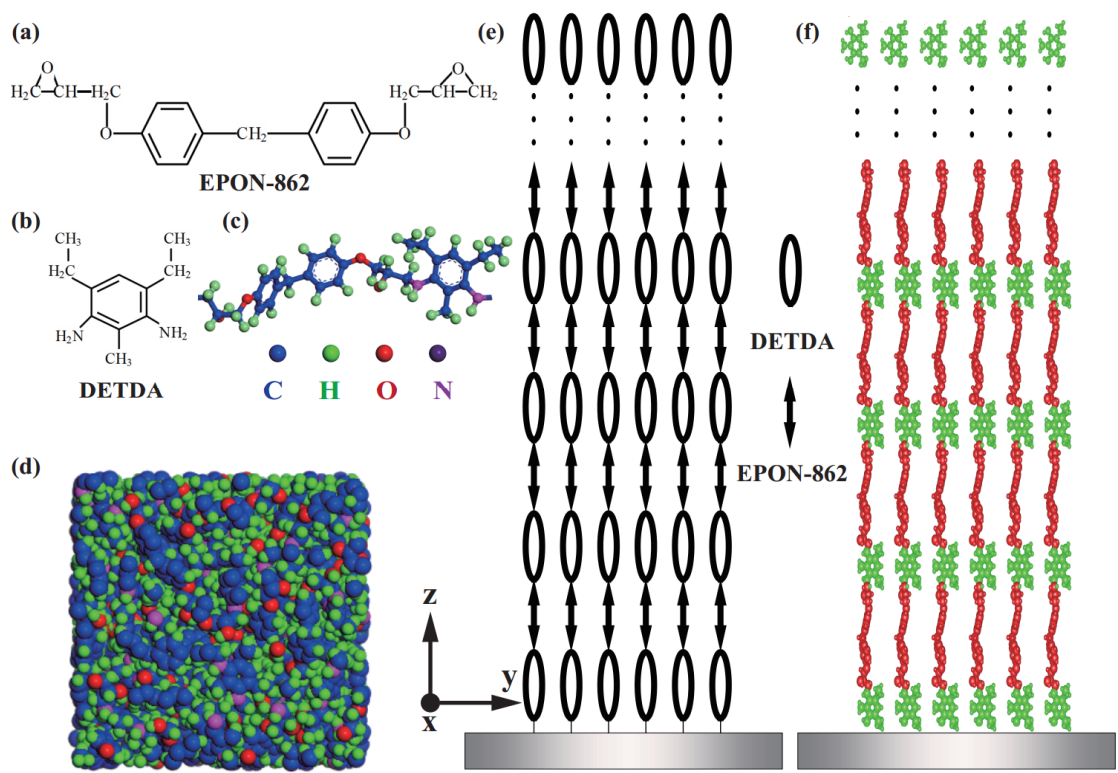

Fig. 1. (a-b) Chemical structures of monomer EPON-862 and curing agent DETDA. (c) One cross-linked segment of epoxy resin, consisting of carbon (blue), hydrogen (green), oxygen (red) and nitrogen (purple). (d) The relaxed cross-linked epoxy resin. (e) An illustration of MLD method. (f) An illustration of PLER structure.

# Results and Discussions

We first calculated the thermal conductivity of CLER with 50% degree of crosslinking and the results are depicted in Fig. 2(a) as an illustration of the EMD method. HCACF fluctuates dramatically for 1 ps, which means the strong reflection of heat current, and then decays to zero rapidly within 3ps for the sake of strong phonon scatterings in CLER, thus giving rise to the quick convergence of integral, i.e., thermal conductivity value. In practical usage of CLER, degree of crosslinking raises with the increase of curing reaction time, so CLER could have different degrees of crosslinking [39]. Therefore, we further investigated the thermal conductivity of amorphous CLER with respect to degree of crosslinking. As Fig. 2(b) shows, the value is in the range of 0.32 - 0.35 $Wm^{-1}K^{-1}$. It is comparable to the previous simulation [12] and experiment results [40], which validates our calculations results about epoxy resins. We can see that degree of crosslinking has little impacts on thermal conductivity of CLER. It is quite different from polyethylene [41] whose thermal conductivity increases with degree of crosslinking. This is attributed to the different contributions of heat transport in these two kinds of cross-linked polymers. Cross-linking bridges the separated molecules in polymers and thus paves the way for heat transport by covalent bonding interactions while it almost has no influences on heat transport by nonbonding interactions. For linear polymer chains, bonding interactions are main contributors [41], so crosslinking further improves the heat transport. However for epoxy resin, it has been verified that nonbonding interactions are the most predominant among different interactions contributions to thermal conductivity [12]. Accordingly, crosslinking has little effects on heat transport in epoxy resin.

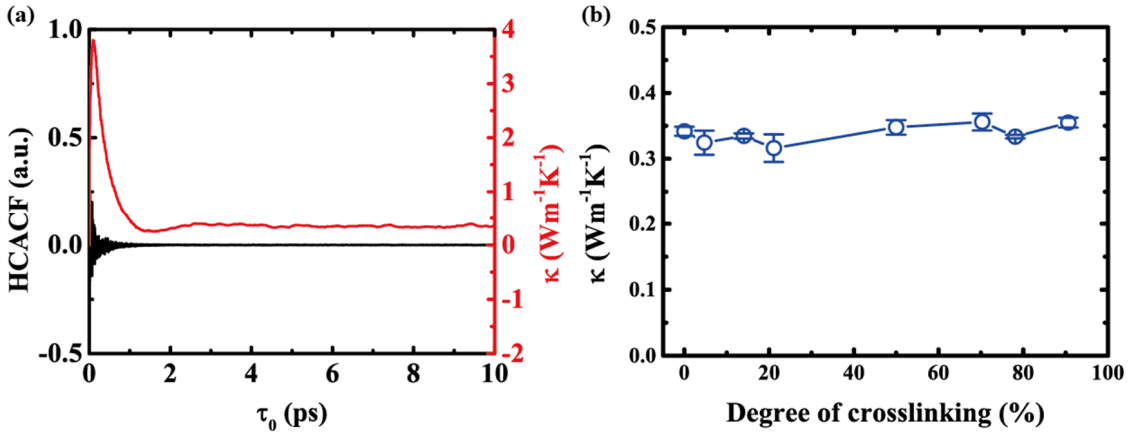

Fig. 2. (a) Normalized heat current autocorrelation function (black line) and thermal conductivity (red line) of cross-linked epoxy resin with 50% degree of crosslinking. (b) Thermal conductivity of cross-linked network epoxy resin as a function of degree of crosslinking.

To show the enhancement on thermal conductivity by bottom-up parallel linking, we performed further simulation on bulk PLER. Due to the well-known simulation size effects of EMD, we performed convergence test with different simulation domain sizes. The thermal conductivity values of systems with 1, 2, 4, 8 segments along chain are calculated and the results are shown in Fig. 3 (a). It can be seen that the simulation domain size effect is not significant for bulk PLER. We use 8 segments as the length of simulation system in the subsequent simulations for bulk PLER. The converged thermal conductivity along intra-chain (z axis) direction ($\kappa_z$) at 300K is 0.80 $Wm^{-1}K^{-1}$, which is twofold higher than that of CLER. Our results are comparable with the thermal conductivity of highly ordered liquid-crystalline epoxy resins as Koda *et al.* [26] reported. Unlike common liquid epoxy resin, the relaxed PLER owns long-range ordered structure, as shown in the inset of Fig. 3(a). Besides, there is no bonding interactions between chains, which can further decrease the intra-chain scatterings. It is noteworthy that the thermal conductivity value can be even further enhanced if monomer and curing agent with stronger backbone are employed in PLER. The thermal conductivities along inter-chain (x and y axes) directions ($\kappa_x$ and $\kappa_y$) are ~ 0.25 Wm$^{-1}$K$^{-1}$, which are close to that of CLER. This is to be expected because the nonbonding interactions are the main contributors for heat transport along x and y directions, while the bonding interactions dominate heat transport along z direction. It can be seen that PLER owns strong anisotropy due to semi-crystal structure (the inset of Fig. 3 (a)), which is much more ordered than amorphous bulk CLER in Fig. 1 (d). However, the thermal conductivity value of PLER is still much smaller than that of bulk aligned PE [18, 42-44] . Previous studies suggested that the thermal conductivities of nanowires and nanoribbons can be decreased with the increasing number of kinks and folds [45-47]. There is also buckling in chains of the relaxed bulk PLER, which indicates that semi-crystalline structure still owns low degree of crystallinity and it can induce intra-chain phonon scatterings, which is also observed in other bulk aligned buckled polymers [48].

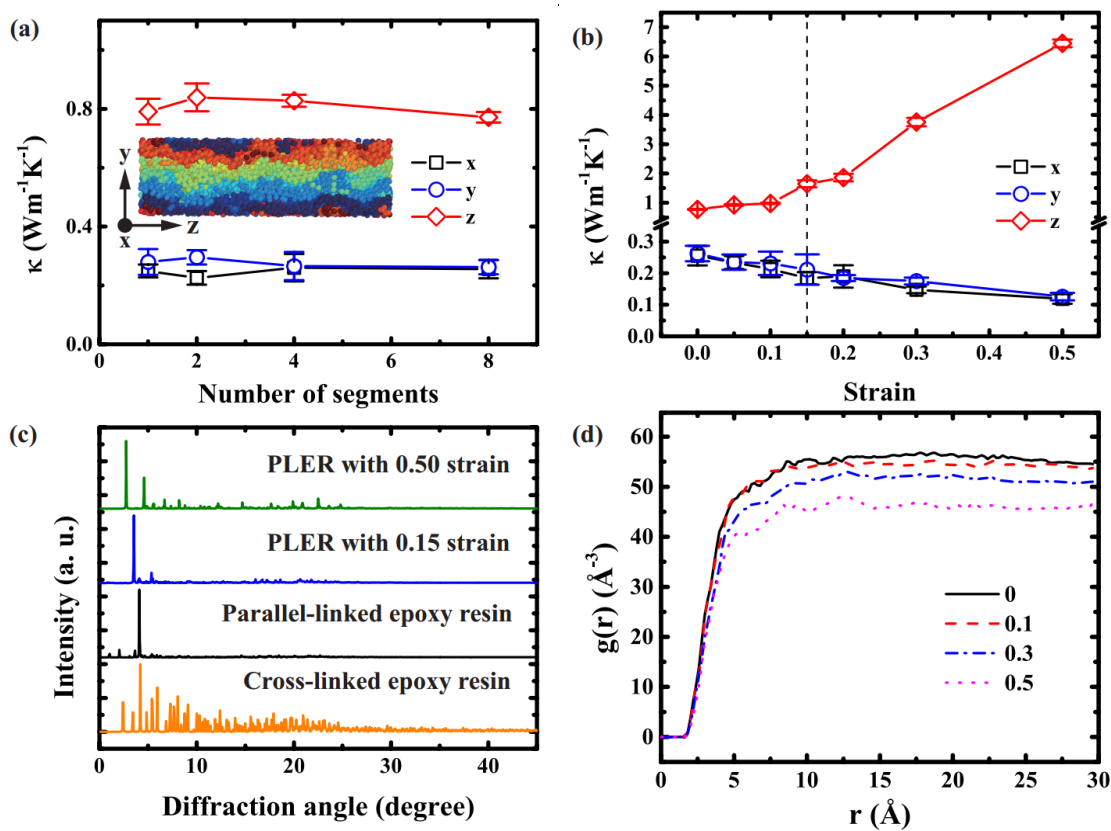

Fig. 3. (a) Thermal conductivity of PLER as a function of number of segments in the chain direction. The inset is a side-view of the relaxed bulk PLER (16 chains are allocated with different colors). (b) The thermal conductivities of stretched PLER with respect to different strains. (c) XRD patterns of different epoxy resin structures. (d) The radial distribution function g(r) as a function of radius r of PLER with different strains. The dash line and dot dash line are the radiuses where peaks in g(r) of PLER without strain and with 0.15 strain locate, respectively.

For further improvement of thermal conductivity, we studied the thermal conductivities of bulk PLER with uniaxial tensile strains [49]. Figure 3(b) shows the strain dependence of thermal conductivity along different directions. It can be seen that κ$_z$ increases dramatically with tensile strain. When the strain is smaller than 0.15, κ$_z$ increases slowly with strain. Then it increases significantly for larger strain. The value of κ$_z$ reaches 6.45 Wm$^{-1}$K$^{-1}$ with 0.50 strain, which is around 20 times larger than that of the bulk CLER. However, κ$_x$ and κ$_y$ decrease from 0.25 Wm$^{-1}$K$^{-1}$ to 0.12 Wm$^{-1}$K$^{-1}$. Similarly, Zeng *et al.* [27] found the thermal conductivity of fibrous epoxy on the through-plane decreases with the strain rising. We did not compute the cases with even larger strain for the confinement of computational ability for the well-known size effects of EMD method.

In order to quantitatively describe the morphologies of CLER, pristine PLER and stretched PLER, we compared the X-ray diffraction (XRD) patterns of three different structures, as it can reflect the crystallinity of structure. The XRD intensity is calculated based on a mesh of reciprocal lattice nodes defined by entire simulation domain using a simulated radiation of wavelength lambda [34, 50]. As presented in Fig. 3(c), the XRD patterns show significant differences among these three structures. As we can see, stretched PLER shows several discrete peaks, and PLER without strain has fewer peaks. CLER shows many broad continuous peaks. The phenomenon indicates that the stretched PLER has a higher crystallinity of structure compared to the structure without strain, while the latter owns more ordered structure compared to the amorphous cross-linked network. There are less phonon scatterings in more ordered structures. That is why the stretched PLER has the highest thermal conductivity among the three structures. Besides, the XRD pattern of PLER with 0.15 strain is similar to that with 0.50 strain. It elucidates that the crystallinity of PLER with strain larger than 0.15 is higher than the ones with smaller strains which results in the saltation of κ$_z$ at the strain 0.15.

We analyzed the inter-chain radial distribution function (RDF) of stretched PLER to further understand the negative strain dependence of inter-chain thermal conductivity. The heat transfer along inter-chain directions is dominated by nonbonding interactions, which are related to the atom distribution around the reference atom. We choose all the atoms in an isolate chain as the reference atoms of RDF. The distance from a reference atom to the atom in other chains is defined as $R = \sqrt{(x-x_0)^2 + (y-y_0)^2 + (z-z_0)^2}$. The RDF is recorded as $g(r) = \frac{n}{\frac{4}{3}\pi[(r+dr)^3 - r^3]}$, where $r$ is the distance between reference atoms and other atoms, and $n$ is the number of atoms with a distance of $R (r < R < r + dr)$ to the reference atoms [18], $dr$ is set to be 0.2 Å. The results are shown in Fig. 3 (d). It can be seen $g(r)$ decreases with the increase of strain. This demonstrates that the reduction of density around the reference atoms and then the decrease of atomic interactions on x and y axes because nonbonding interactions are the leading contributors of heat propagation on these two axes. Hence, the interactions on x and y axes would be smaller for stretched PLER with larger tensile strains and then κ$_x$ and κ$_y$ would be lower. The differences of $g(r)$ are more significant when the strain is larger than 0.3. Corresponding, κ$_x$ and κ$_y$ are decrease dramatically.

To gain insights into how the inter-chain nonbonding interactions impact on the thermal conductivity of PLER, the single chain structure (Fig. 4 (a)) was investigated which can avoid such inter-chain phonon scatterings, and help us obtain the upper limit thermal conductivity of epoxy resin. In this study, a sufficiently large vacuum space was introduced artificially to ensure that the single chain is isolate and there are no other chains interacting with it. The relaxed chain length obtained from the bulk PLER is used for isolated single chain at the same temperature. We deformed the foregoing relaxed single chain and bulk structure with corresponding strains and we could get the stretched structures. The thermal conductivity of unstretched PLER single chain structure is shown in Fig. 4 (b). The converged value is ~ 1.13 Wm$^{-1}$K$^{-1}$ on z axis, which is about threefold higher than that of amorphous cross-linked network and larger than that of bulk PLER. The converged length of single chain structure is much larger than the bulk PLER, which can also verify that the absence of inter-chain interactions could reduce the phonon scatterings and facilitate the propagation of the long-wavelength phonons. We also calculated the thermal conductivity of stretched PLER single chain. The tensile strain increases from 0 to 0.80 and it can have strong effects on the morphology of the single chain and greatly affect $\kappa_z$ of PLER single chain, as shown in Fig. 4 (a). The values increase by 30 times, from ~ 1.13 Wm$^{-1}$K$^{-1}$ to ~ 33.82 Wm$^{-1}$K$^{-1}$. The variation trend of thermal conductivity with strain is analogous to that of stretched bulk PLER: The thermal conductivity increases slowly with small strain and then increases linearly when the strain is higher than 0.15. As shown in Fig. 4(a), the morphology of single chain with 0.15 strain is straighter than that of unstretched structure and this phenomenon is more significant in the structure with 0.30 strain. The thermal conductivity still keeps the increasing trend with strain raising. This implies the thermal conductivity of bulk PLER could be further enhanced through decreasing both intra-chain and inter-chain phonon scatterings.

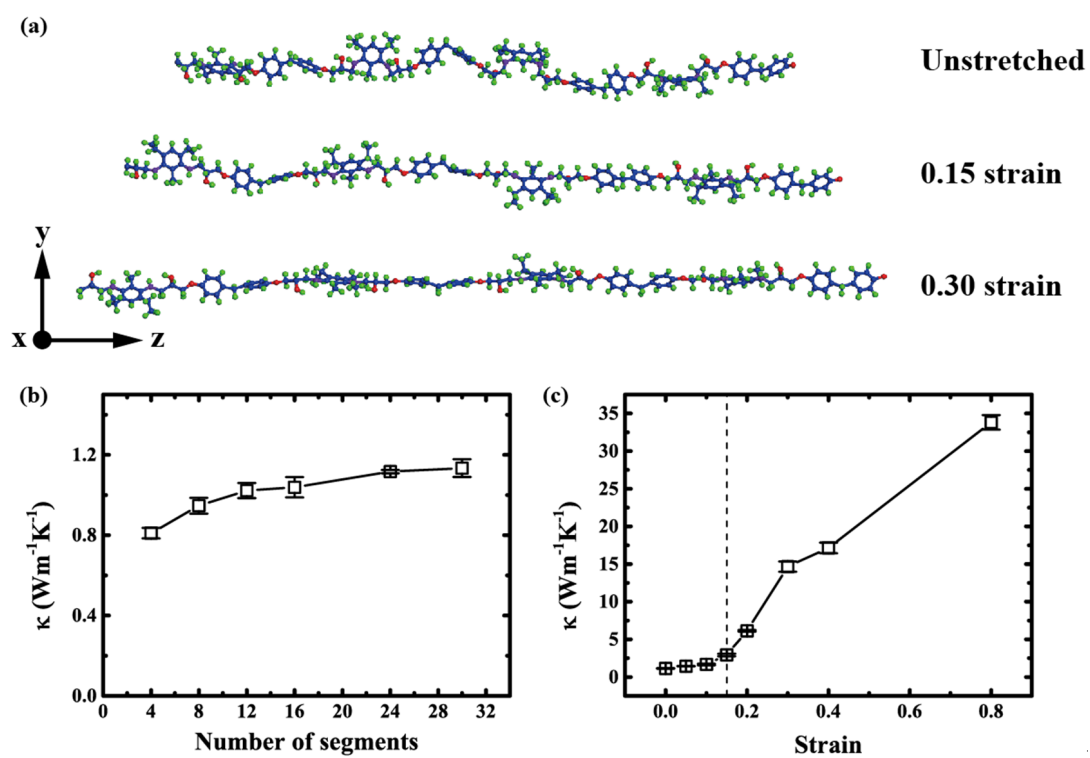

Fig. 4. (a) Epoxy resin single chains with 0.15 and 0.30 strain. (b) The thermal conductivities of single chain with respect to different number of segments. (c) The thermal conductivities of stretched single chain as a function of strains.

# Conclusions

In summary, we performed equilibrium molecular dynamics simulations to investigate the thermal conductivities of cross-linked and parallel-linked epoxy resins as well as epoxy resin single chain. Our results show that the thermal conductivity of cross-linked epoxy resin is in the range of 0.32 - 0.35 $Wm^{-1}K^{-1}$ and the degree of crosslinking has little effects because the nonbonding interactions are the main contributors in heat transport. We proposed a bottom-up parallel-linking method, which is shown to be efficient and the along-chain thermal conductivity of bulk parallel-linked epoxy resin structure can be enhanced more than twofold compared to the cross-linked network structure due to the much more ordered morphology. The thermal conductivity can be boosted with the increasing strain, and reaches a value as high as 6.45 $Wm^{-1}K^{-1}$ with 0.50 strain. The thermal conductivity firstly increases slightly and then increases significantly with the strain raising. X-ray diffraction pattern shows that deformed parallel-linked epoxy resin owns perfect crystalline structure, which explains the surprising high thermal conductivity. The thermal conductivity on x and y axes would decrease with the increase of strain for degraded nonbonding interactions. For parallel-linked single chain, the thermal conductivity is about threefold higher than that of cross-linked network and the value can be enhanced greatly by 30 times, reaching as high as ~ 33.82 $Wm^{-1}K^{-1}$ when strain reaches 0.80. The stretched structures own higher thermal conductivities than that of unstreched one due to more ordered morphology and the suppression of both inter-chain and intra-chain phonon scatterings. The results may provide a meritorious guide on the design and fabrication of epoxy resin with high thermal conductivity.

# Acknowledgements

This work was supported by the National Natural Science Foundation of China No. 51676121 (HB), 51576076 (NY) and 51711540031 (NY), and the Hubei Provincial Natural Science Foundation of China No. 2017CFA046 (NY). Simulations were performed with computing resources granted by HPC (π) from Shanghai Jiao Tong University and the National Supercomputing Center in Tianjin (NSCC-TJ).